\documentclass[oldversion]{aa}  
\usepackage{graphicx,amsmath}
\usepackage{amssymb}
\usepackage{color}
\usepackage{natbib}		
\usepackage{url}
\usepackage{multirow}
\usepackage{threeparttable}   
\usepackage{soul}
\bibpunct{(}{)}{;}{a}{}{,} 

\begin{document}

   \title{Refined architecture of the WASP-8 system: a cautionary tale for traditional Rossiter-McLaughlin analysis.}
   				   
   \author{
   V. Bourrier\inst{1},
   H.~M. Cegla\inst{1},
   C. Lovis\inst{1},
   A. Wyttenbach\inst{1}
	}
   
\authorrunning{V.~Bourrier et al.}
\titlerunning{Refined architecture of the WASP-8 system}

\offprints{V.B. (\email{vincent.bourrier@unige.ch})}

\institute{
Observatoire de l'Universit\'e de Gen\`eve, 51 chemin des Maillettes, 1290 Sauverny, Switzerland
}
   
   \date{} 
 
  \abstract
{Probing the trajectory of a transiting planet across the disk of its star through the analysis of its Rossiter-McLaughlin effect can be used to measure the differential rotation of the host star and the true obliquity of the system. Highly misaligned systems could be particularly conducive to these mesurements, which is why we reanalysed the HARPS transit spectra of WASP-8b using the ``Rossiter-McLaughlin effect reloaded" (reloaded RM) technique. This approach allows us to isolate the local stellar CCF emitted by the planet-occulted regions. As a result we identified a $\sim$35\% variation in the local CCF contrast along the transit chord, which might trace a deepening of the stellar lines from the equator to the poles. Whatever its origin, such an effect cannot be detected when analyzing the RV centroids of the disk-integrated CCFs through a traditional velocimetric analysis of the RM effect. Consequently it injected a significant bias into the results obtained by Queloz et al. (2010) for the projected rotational velocity $v_{eq} \sin i_{\star}$ (1.59$\stackrel{+0.08}{_{-0.09}}$\,km\,s$^{-1}$) and the sky-projected obliquity $\lambda$ (-123.0$\stackrel{+3.4}{_{-4.4}}^{\circ}$). Using our technique, we measured these values to be $v_{eq} \sin i_{\star}$ = 1.90$\pm$0.05\,km\,s$^{-1}$ and $\lambda$ = -143.0$\stackrel{+1.6}{_{-1.5}}^{\circ}$. We found no compelling evidence for differential rotation of the star, although there are hints that WASP-8 is pointing away from us with the stellar poles rotating about 25\% slower than the equator. Measurements at higher accuracy during ingress/egress will be required to confirm this result. In contrast to the traditional analysis of the RM effect, the reloaded RM technique directly extracts the local stellar CCFs, allowing us to analyze their shape and to measure their RV centroids, unbiased by variations in their contrast or FWHM.}


   \maketitle

\section{Introduction}
\label{intro} 

\subsection{Analysis of the Rossiter-McLaughlin effect} 

The occultation of a rotating star by a planet removes from the apparent stellar line shape the profile part emitted by the hidden portion of the stellar photosphere. This distortion of the spectral lines, known as the ``Rossiter-McLaughlin (RM) effect" (\citealt{holt1893}; \citealt{rossiter1924}; \citealt{mclaughlin1924}) traces the trajectory of the planet across the surface of the stellar disk, and is thus sensitive to the velocity of the stellar photosphere and to the alignment between the spins of the planetary orbit and the stellar rotation (namely the obliquity). Measuring the spin-orbit alignment is particularly important because it can feed into theories on planetary migration and evolution. To date, the sky-projected obliquity has bean measured for more than 90 planetary systems\footnote{\mbox{the Holt-Rossiter-McLaughlin Encyclopaedia in Oct. 2016:} \mbox{\url{http://www2.mps.mpg.de/homes/heller/}}} (\citealt{albrecht2012}; \citealt{Crida2014}). Most of these measurements have been carried out using Doppler spectroscopy, as the distortion of the stellar lines and their corresponding cross-correlation function (CCF) induces deviations to the Keplerian radial velocity (RV) of the star during the transit that can be fitted with analytical formulae (e.g. \citealt{ohta2005}, \citealt{gimenez2006}, \citealt{hirano2011b}, \citealt{boue2013}). However, because it only fits this RV anomaly and not the full spectral CCF, this technique (hereafter referred to as the velocimetric analysis of the RM effect) can be prone to significant biases (e.g., \citealt{Triaud2009}, \citealt{Hirano2010}, \citealt{Brown2016}). For example, fitting the RV anomaly using (analytical or numerical) models that assume the local stellar profile is a constant Gaussian can potentially inject biases of up to 20-30$^{\circ}$ in the projected obliquity \citep{Cegla2016a}.

In recent years, Doppler tomography has been used to study the alignement and properties of a growing number of planetary systems (eg \citealt{cameron2010a}, \citealt{Gandolfi2012}; \citealt{bourrier2015_koi12}). This technique relies on the decomposition of the observed CCF profile into its different components (i.e., the stellar and instrumental profiles, the limb-darkened rotation profile, and the missing starlight signature caused by the planet). As such, it removes some of the biases of the velocimetric analysis and can provide additional constraints on the planet-to-star radius ratio and the intrinsic stellar line profile. Yet even Doppler tomography assumes a constant, symmetric line profile across the stellar disk, and typically ignores the effects of differential rotation and velocity shifts due to stellar oscillations and granulation. 

This motivated \citet{Cegla2016} to develop a new technique to analyse and model the RM effect (the ``Rossiter-McLaughlin effect Reloaded" technique, hereafter reloaded RM). By using the planet as a probe, the reloaded RM technique isolates the local CCFs from the regions successively occulted during the transit, with no particular assumptions on the shape of the intrinsic stellar photospheric lines. Many properties can be derived from these local CCFs, including the RV of the stellar surface along the transit chord. In turn, these RVs can be fitted with a simple model to derive the sky-projected obliquity and the stellar rotational velocity. Furthermore, because it accounts for differential rotation, the reloaded RM technique can be sensitive to the stellar latitudes occulted by the planet. This can be used to lift the degeneracy between the equatorial velocity and the stellar inclination, therefore yielding the value of the true 3D obliquity.

\citet{Cegla2016} benchmarked the reloaded RM technique on the HD189733b system, and were able to constrain the differential rotation of the host star. However, HD189733b transits its star at a relatively low latitude ($\sim$40$^\circ$), along a nearly aligned orbit, and the local CCFs measured along the transit chord showed very similar properties. This motivated our search for a planet occulting more diverse regions of its host star, making it even more conducive to the application of the reloaded RM technique. \\

\subsection{The WASP-8 system} 

Among all systems with known obliquities, we identified WASP-8 as one of the most suitable targets. It is a binary stellar system 87\,pc away from Earth, with the A component a bright G-type star (V=9.8) hosting two exoplanets WASP-8b and WASP-8c, and the B component a faint M dwarf. The two stars have a common high proper motion (111 mas/year), and their angular separation (4.495'') sets a minimum separation of 440\,au (\citealt{evans2016}). WASP-8b was first detected by \citet{queloz2010} through transit observations with the WASP-south telescope, and then followed in radial velocity with the Coralie spectrograph. \citet{queloz2010} further reported a linear trend in this radial velocity follow-up, which was later revealed by new Keck HIRES measurements as the reflex motion from the outer companion WASP-8c (\citealt{knutson2014_HJ}; P = 4339$\stackrel{+850}{_{-390}}$\,days, M\,sin\,i = 9.5$\stackrel{+2.7}{_{-1.1}}$\,M$_\mathrm{jup}$ assuming a circular orbit). WASP-8b is a Jupiter-size planet (1.04\,R$_\mathrm{jup}$; \citealt{queloz2010}) with a mass of 2.24\,M$_\mathrm{jup}$ (\citealt{knutson2014_HJ}). It is on an eccentric ($e$=0.3), 8.16\,days orbit. Its spectroscopic transit was measured with the HARPS spectrograph (\citealt{queloz2010}), revealing the radial velocity anomaly arising from the RM effect caused by a retrograde, highly misaligned orbit. 

The availability of archival HARPS transit spectra for WASP-8b, their high SNR ratio, and the large planet-to-star surface ratio ($\delta\sim$1.3\%) makes this sytem a target of choice for the application of the reloaded RM technique. From its large impact parameter ($b$=0.6) and large sky-projected obliquity ($| \lambda | > $ 90$^{\circ}$) we further expect to probe very different regions of the host star surface, as the transit chord roughly goes from the sky-projected spin axis at ingress to the stellar equator at egress. Finally, with an age of 4\,Gy old and a G\,6-type, WASP-8 is also a good candidate to search for differential rotation similar to our Sun.

In Sect.~\ref{sec:data_red}, we describe the spectroscopic transit dataset of WASP-8b and how it is processed to extract the CCFs of the local stellar photosphere. In Sect.~\ref{sec:rv_field} we present our reloaded RM model for the WASP-8 system, and the measurements derived for the system architecture and stellar surface properties. Sect.~\ref{sec:shape_ccf} discusses changes in the local CCF shape across the stellar disc and their interpretation based on simple star-planet simulations. In Sect.~\ref{sec:pitfalls} we compare different techniques used to analyse the RM effect. We conclude and discuss the significance of our results in Sect.~\ref{sec:conclu}. Properties of the WASP-8 system that were used in our analysis as fixed parameters are given in Table.~\ref{tab:fix}.

\section{Observations and data reduction}
\label{sec:data_red} 

\subsection{HARPS transit spectra}

We used the transit observations of WASP-8b (\citealt{queloz2010}) obtained with the HARPS echelle spectrograph mounted on the ESO 3.6 m telescope in La Silla, in Chile. 75 exposures of 300\,s duration were secured during the single night of October 4, 2008, covering the full duration of the transit (47 exposures) and providing a good baseline for the unocculted host star both before (7 exposures) and after (20 exposures) the transit. During the measurement of the transit sequence a significant change in telescope focus occurred at orbital phase -0.0014, leading to a better accuracy and lower dispersion on the parameters derived from the reloaded RM analysis (e.g., Fig.~\ref{fig:RV_series}). Observations were reduced with the version 3.5 of the HARPS Data Reduction Software, yielding spectra with resolution 115000 covering the region between 380 nm and 690 nm. The reduced spectra were passed through an order-by-order cross-correlation with a G2-type mask function, weighted by the depth of the lines. Since the output CCFs are oversampled with a step of 0.25\,km\,s$^{-1}$, for a pixel width of about 0.8\,km\,s$^{-1}$, we kept one in four points in all CCFs prior to their analysis. The CCFs are defined in the Solar System barycenter rest frame.

\begin{table}[b!]
\caption[]{Fixed parameters for the WASP-8 system}
\centering
\begin{threeparttable}
\begin{tabular}{c|c|c}
    \hline
    \hline
      Parameters & Value & Reference \\
    \hline  
	$R_{\star}$ & 0.945$\stackrel{+0.051}{_{-0.036}}$ R$_{\odot}$& \citealt{queloz2010}\\    
	$R_{\star}/a$ & 0.0549  $\pm$ 0.0024 & \citealt{queloz2010} \\
	$u_1$ & 0.516 & \citealt{queloz2010}$^{a}$ \\
	$u_2$ & 0.224   & \citealt{queloz2010}$^{a}$ \\
	T$_\mathrm{eff}$ & 5600$\pm$80\,K & \citealt{queloz2010}\\    
	log $g$ & 4.5 $\pm$ 0.1 & \citealt{queloz2010}\\	
	$T_{0}$ & 2454679.33394$\stackrel{+0.00050}{_{-0.00043}}$ d & \citealt{queloz2010} \\    
	$P$ & 8.158715 $\pm$ 1.6$\times$10$^{-5}$ d & \citealt{queloz2010}\\    
	$i_p$ & 88.55$\stackrel{+0.15}{_{-0.17}}$ $^{\circ}$ & \citealt{queloz2010} \\    
	$(R_p/R_{\star})^2$ & 0.01276  $\pm$ 0.00033  & \citealt{queloz2010}\\
	$K$ & 221.1 $\pm$ 1.2 m\,s$^{-1}$ & \citealt{knutson2014_HJ} \\
	$e$ & 0.304$\pm$0.004     &  \citealt{knutson2014_HJ} \\
	$\omega$ & 274.22$\pm$0.084  $^{\circ}$ & \citealt{knutson2014_HJ}  \\
    \hline
  \end{tabular}
  \begin{tablenotes}[para,flushleft]
  Notes: $^{(a)}$ The limb-darkening coefficients were inferred from \citealt{queloz2010} parameters for the host star. See text.\\

  \end{tablenotes}
  \end{threeparttable}
\label{tab:fix}
\end{table}

\subsection{The reloaded RM technique}
\label{sec:reloaded RM}
\begin{figure}
\centering
\includegraphics[trim=0cm 2.3cm 0.5cm 0cm,clip=true,width=\columnwidth]{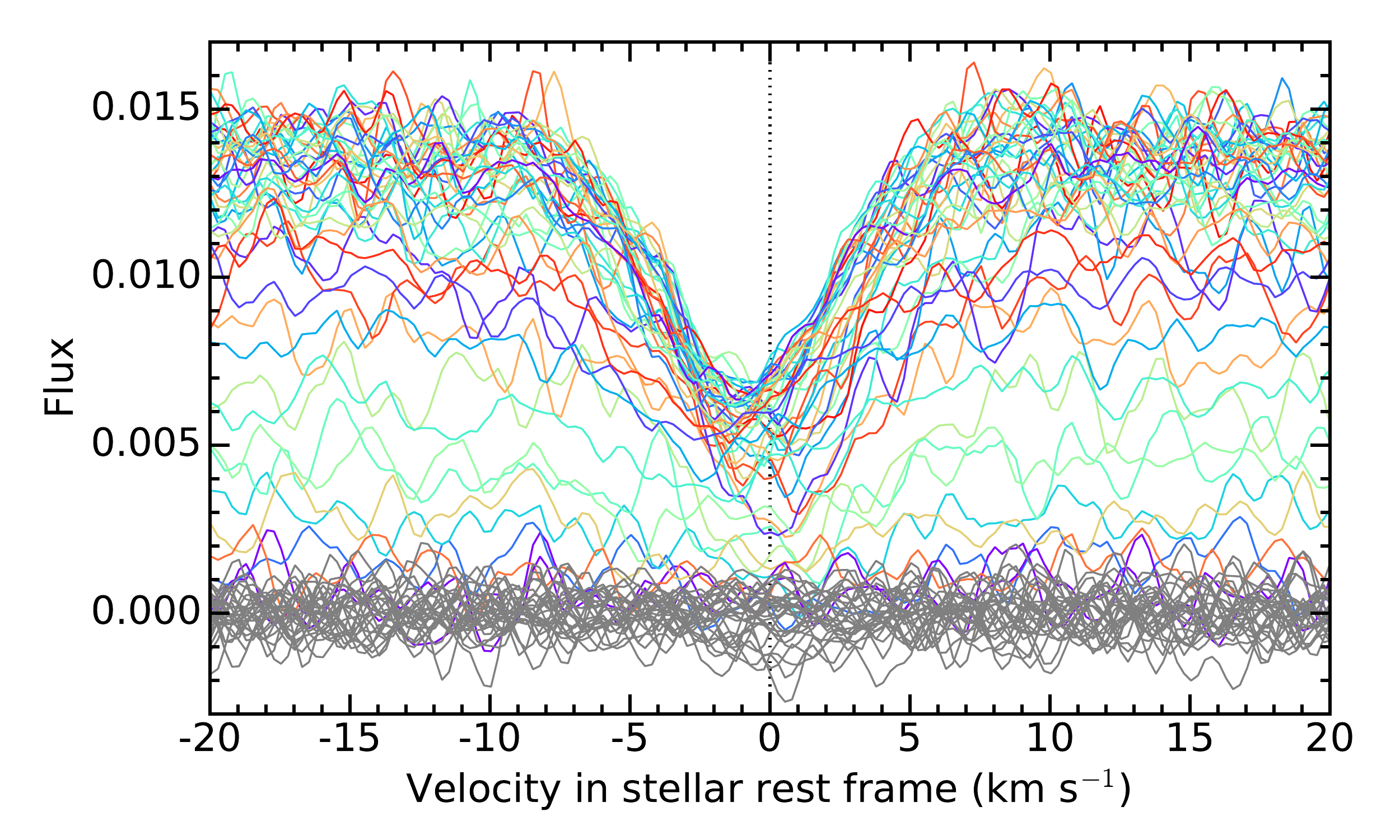}
\includegraphics[trim=0cm 0.5cm 0.5cm 7.2cm,clip=true,width=\columnwidth]{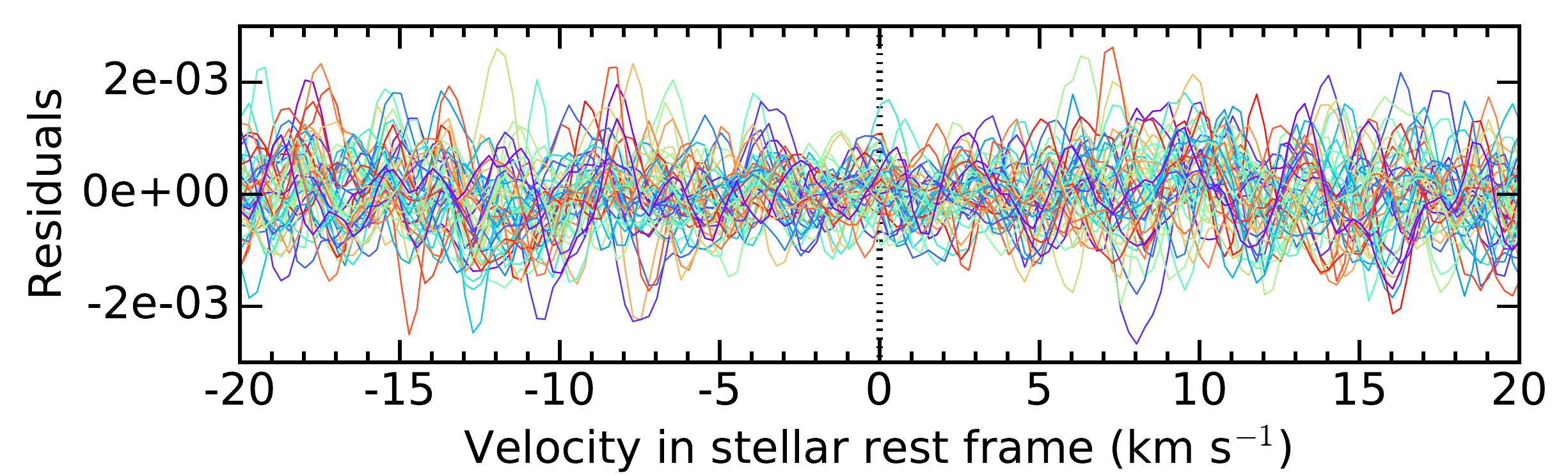}
\caption[]{\textit{Top panel:} Local CCF$_\mathrm{loc}$ profiles (with colors, chosen for viewing ease) and residuals between the out-of-transit CCF$_\mathrm{DI}$ and the master-out (grey color). \textit{Bottom panel:} Residuals between the CCF$_\mathrm{loc}$ profiles and their Gaussian fits. The dotted black line marks the star rest velocity.}
\label{fig:CCF_locales}
\end{figure}

We used the reloaded RM technique developed by \citet{Cegla2016} to isolate the starlight from the regions occulted by the planet during its transit. Hereafter we distinguish between the local CCFs (CCF$_\mathrm{loc}$), which correspond to the light emitted by the planet-occulted regions, and the disk-integrated CCFs (CCF$_\mathrm{DI}$), which corresponds to the light emitted by the entire stellar disk (minus the planet-occulted regions during the transit). First, the CCF$_\mathrm{DI}$ obtained from the HARPS spectra were shifted in velocity space, correcting for the star's Keplerian motion induced by WASP-8b (using the orbital properties in Table~\ref{tab:fix}). We do not account for WASP-8c, as it has a long period (4339\,d; \citealt{knutson2014_HJ}) and induces a negligible reflex motion on the host star during the $\sim$9\,h span of the observations ($\sim$8\,cm\,s$^{-1}$). We then separated the CCF$_\mathrm{DI}$ secured outside of the transit and those within, using the ephemeris given in Table~\ref{tab:fix}. CCF$_\mathrm{DI}$ outside of the transit were co-added to build a single ``master-out'' CCF$_\mathrm{DI}$, whose continuum was normalized to unity. The continuum is defined as regions farther than 10\,km\,s$^{-1}$ from the star rest velocity, that corresponds to $\gamma_{\star}$ in the barycentric frame. Fitting the master-out with a Gaussian profile, we found its centroid at $\gamma_{\star}$  = -1537.5$\pm$5.5\,m\,s$^{-1}$. The first spectrum of the night was excluded from our analysis because the residuals between its out-of-transit CCF$_\mathrm{DI}$ and the master-out revealed a spurious signature, probably caused by twilight illumination. Using the value obtained for $\gamma_{\star}$, all CCF$_\mathrm{DI}$ were then shifted to the stellar rest frame. Since the HARPS observations are not calibrated photometrically, each in-transit CCF$_\mathrm{DI}$ has to be continuum-scaled to reflect the planetary disk absorption. To do this, we used the photometric light curve fitted to the WASP-south photometry by \citet{queloz2010}. Since the quadratic limb-darkening coefficients were not available in their paper, we retrieved them using the EXOFAST calculator\footnote{\url{http://astroutils.astronomy.ohio-state.edu/exofast/limbdark.shtml}} (\citealt{Eastman2016}) with \citet{queloz2010} values\footnote{Note that we used the \citet{queloz2010} T$_\mathrm{eff}$ value for consistency, but a higher temperature was obtained by \citet{evans2016}.} for T$_\mathrm{eff}$, log $g$ and [Fe/H]. We note that our light curve, calculated with the \textit{batman} package (\citealt{kreidberg2015}), matches well the one in \citet{queloz2010}. Finally the local CCF$_\mathrm{loc}$ associated to the planet-occulted regions were retrieved by subtracting the scaled in-transit CCF$_\mathrm{DI}$ from the master-out (Fig.~\ref{fig:CCF_locales}). 

The CCF$_\mathrm{loc}$ resolve spectrally and spatially the photosphere of the star along the transit chord. They have a lower continuum flux near the limb of the stellar disk due to the assumed limb darkening, and the partial occultation from the planetary disk during ingress/egress (Fig.~\ref{fig:CCF_locales}). We analyzed their shape and its possible variations across the stellar disk by fitting independent Gaussian profiles to each CCF$_\mathrm{loc}$, and deriving their RV centroid, contrast and full width at half maximum (FWHM). The flux errors assigned to the CCF$_\mathrm{loc}$ were derived from the standard deviation in their continuum flux, and the uncertainties on the derived parameters correspond\st{s} to the 1\,$\sigma$ statistical errors from the Levenberg-Marquardt least-squares minimisation. As can be seen in Fig.~\ref{fig:CCF_locales}, the residuals between the CCF$_\mathrm{loc}$ and their fit yield a low dispersion and no particular features, showing that the assumption of a Gaussian profile for the CCF$_\mathrm{loc}$ is satisfactory. Hereafter we exclude from our analysis exposures for which the CCF$_\mathrm{loc}$ was not detected, or detected with a contrast lower than three times the dispersion of the residuals in the continuum. This concerns the exposures at orbital phases lower than -0.01 (two at ingress) and higher than 0.0095 (three at egress; see the upper panel in Fig.~\ref{fig:RV_series}).

\section{Analysis of the stellar surface velocity field}
\label{sec:rv_field}

\subsection{Method}
\label{sec:method}
\begin{figure}
\centering
\includegraphics[trim=1cm 1.6cm 0cm 0cm,clip=true,width=\columnwidth]{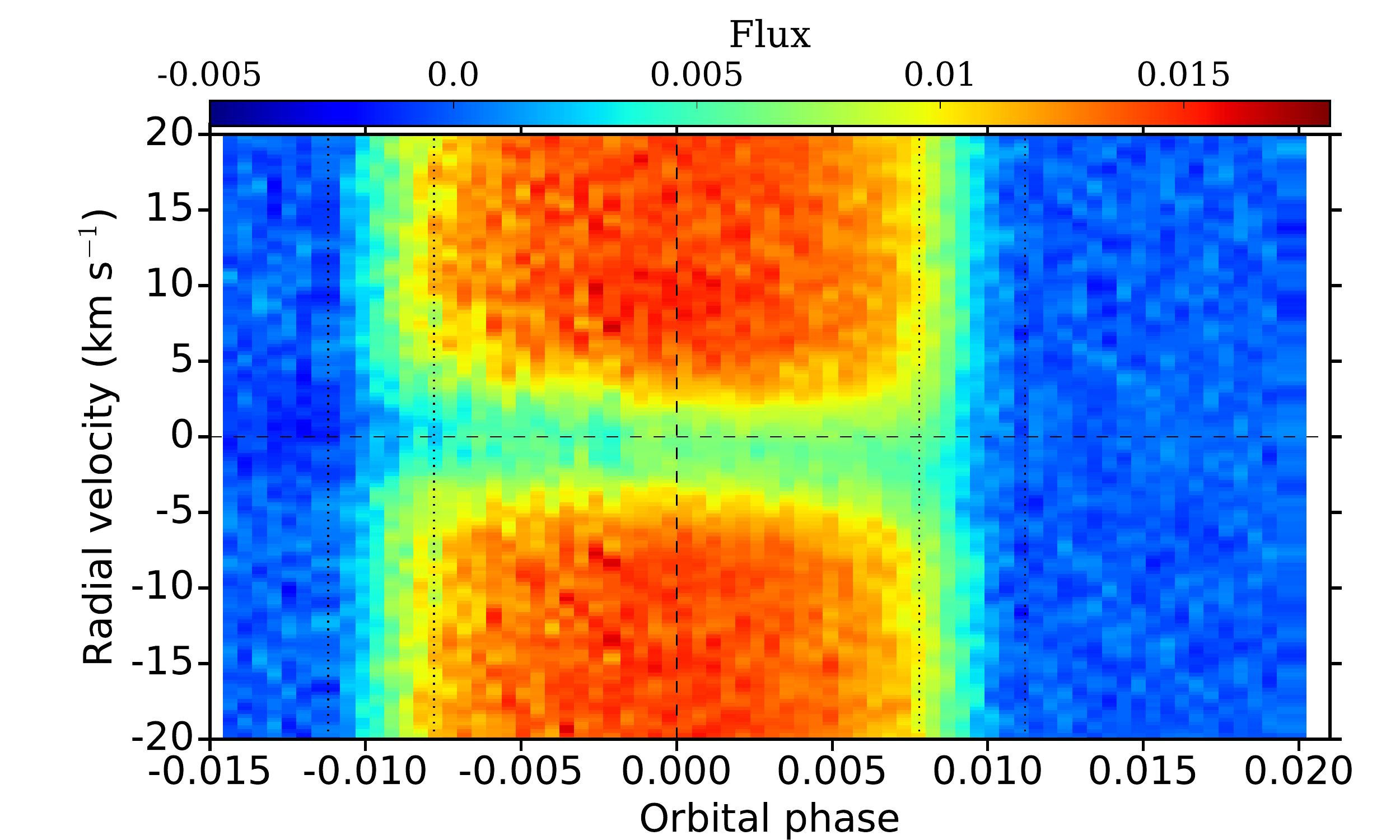}
\includegraphics[trim=1cm 1.8cm 0cm 3cm,clip=true,width=\columnwidth]{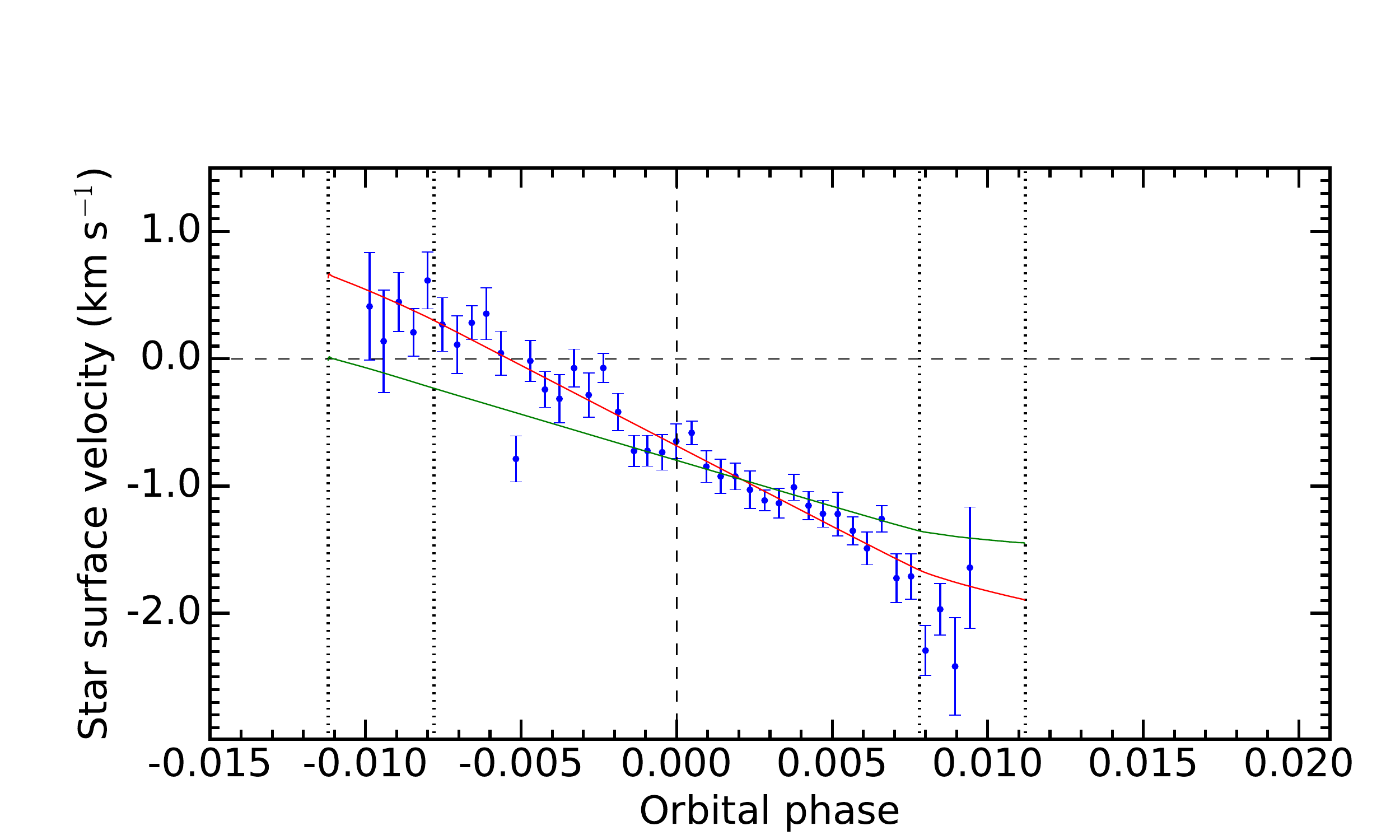}
\includegraphics[trim=1cm 0cm 0cm 9cm,clip=true,width=\columnwidth]{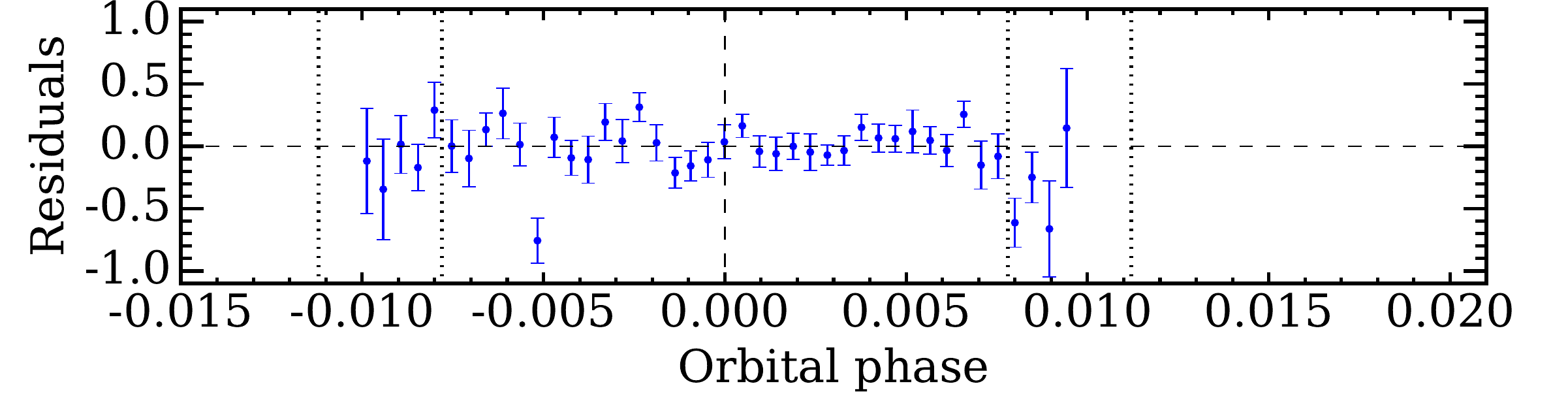}
\caption[]{\textit{Upper panel:} Map of the CCF$_\mathrm{loc}$ series, as a function of orbital phase (in abscissa) and radial velocity in the stellar rest frame (in ordinate). Colors indicate flux values. Vertical and horizontal dashed black line indicate respectively the mid-transit time and stellar rest velocity. The four vertical dashed black lines show the times of the contacts. Note the in-transit CCF$_\mathrm{loc}$ correspond to the average stellar line profiles from the regions occulted by WASP-8b across the stellar disk. \textit{Middle panel:} RVs of the stellar surface regions occulted by the planet (blue points). The green curve is the RV model obtained for the obliquity and rotational velocity in \citet{queloz2010}. The red curve corresponds to our best-fit for the solid-body rotation case. \textit{Lower panel:} Residuals between the measured surface RV and our best-fit shown in the middle panel. }
\label{fig:RV_series}
\end{figure}

The RV of the planet-occulted regions are displayed in Fig.~\ref{fig:RV_series}. They were interpreted using the model developed by \citet{Cegla2016}, which consists of a single formula that depends on the planet's stellar disk position, its sky-projected obliquity ($\lambda$), the stellar limb-darkening, the stellar inclination ($i_{*}$), the star's equatorial rotational velocity ($v_\mathrm{eq}$), its differential rotation, and convective velocities. Our definitions for the coordinate system and angle conventions are the same as in \citet{Cegla2016} (see their Figure 3). The stellar inclination (in 0--180$^{\circ}$) is the angle counted positive from the line-of-sight toward the star spin axis, while $\lambda$ (in  -180--180$^{\circ}$) is the angle counted positive from the star spin axis toward the orbital plane normal, and projected in the plane of sky. Different scenarios for the reloaded RM model are used in Sect.~\ref{sec:res_velfield}, depending on whether or not we account for differential rotation and convective effects. In each case, the theoretical RV of a planet-occulted region is calculated as the brightness-weighted average of the model RVs sampled over a square grid with resolution about 7$\times$10$^{-3}$ the stellar radius (equal to R$_{p}$/31). Contributions from cells that do not lie beneath the planet are excluded. Furthermore, because a transiting planet can move significantly during the duration of an exposure, the occulted regions should not be modeled by a simple disk. To account for this blur, we oversample the chord between the planet position at the beginning and end of each exposure with a resolution $R_{p}$/20. The final theoretical RV, comparable to the measured RV, is the brightness-weighted average of all RVs from the planet grid at every oversampled position. We found, however, that the blur is negligible during the 300\,s long exposures of WASP-8b. 

We fit the reloaded RM model parameters using the Metropolis-Hasting Markov chain Monte Carlo (MCMC) algorithm detailed in \citet{bourrier2015_koi12} (see also \citealt{Tegmark2004}, \citealt{Ford2006}). An adaptive principal component analysis is applied so that step jumps take place in an uncorrelated space, allowing us to better sample the posterior distributions in the eventuality of non-linear correlations between parameters (\citealt{diaz2014}). Jump parameters are described in the following sections depending on the scenario investigated. Except when specified, uniform priors were used on all parameters. The system is analyzed with multiple chains, started at random points in the parameter space, and with a jump size adjusted to get an acceptance rate of $\sim$25\%. We check that all chains converge to the same solution, before thinning them using the maximum correlation length of all parameters. Finally, the thinned chains are merged so that the posterior distributions contain a sufficient number of independent samples. The best-fit values for the model parameters are set to the medians of the posterior probability distributions, and their 1$\sigma$ uncertainties are evaluated by taking limits at 34.15\% on either side of the median. We compare the results from the different reloaded RM models by calculating the Bayesian information criterion for the best-fit (there are 42 fitted RV points). 

\subsection{Stellar rotation and obliquity results} 
\label{sec:res_velfield}
Before the application of the reloaded RM model, information can be retrieved from a visual analysis of the local RVs series (Fig.~\ref{fig:RV_series}). The measured velocities decrease with orbital phase, going from positive to negative values, which shows that WASP-8b is on a retrograde orbit crossing first the redshifted regions of the stellar disk (moving away from the observer) and then its blueshifted regions. Furthermore, the strong asymmetry with respect to the phase / velocity origin indicates a significantly misaligned orbit with more time spent occulting the blueshifted regions.

\subsubsection{Solid-body scenario} 
To go further in the analysis of the system, we first applied the reloaded RM model assuming a solid-body (SB) rotation for the star. In that case, the absence of differential rotation leads to a degeneracy between $v_\mathrm{eq}$ and $sin\,i_{*}$ that prevents us from determining the exact stellar latitudes transited by the planet. The two jump parameters of the MCMC are thus the sky-projected obliquity $\lambda$ and the projected rotational velocity $v_\mathrm{eq}\,sin\,i_{*}$. Their posterior probability distributions are shown in Fig.~\ref{fig:mcmc_solidbody}, along with their marginalized 1D distributions that are well represented by Gaussians and allowed us to derive tight constraints on the inferred best-fit values (Table~\ref{tab:MCMC_results}). The RV model obtained for these values provides a very good fit to the data (Fig.~\ref{fig:RV_series}), and the corresponding system architecture is displayed in Fig.~\ref{fig:disque}. We note that the residual at phase near -0.005 is significantly larger than its uncertainty. It is possible that the planet occulted a stellar spot during this exposure. However the planet would have had to be grazing the spot to yield such an isolated variation in phase, and the amplitude of the variation is quite large to be caused by a spot ($\sim$800\,m\,s$^{-1}$). Furthermore, the exposure does not show similar deviations to the contrast and FWHM time-series (Fig.~\ref{fig:ctrst_sig}). Thus, the outlying RV value could rather be a statistical deviation and we did not exclude the corresponding exposure from our analysis.

We then accounted for centre-to-limb convective variations, including in the model a polynomial law in $\mu = cos(\theta)$ ($\theta$ is the center-to-limb angle) for the convective velocities (CV). It is not possible to detect the constant component of the CV because it is degenerate with the true stellar rest velocity in the $\gamma_{\star}$ measurement of the master-out $CCF_{DI}$, which was subtracted from the data. The model brightness-weighted CV integrated over the stellar disc must therefore equal zero, which implies that the zeroth order coefficient of the polynomial law ($c_\mathrm{0}$) can only be determined by the higher-order coefficients, which are added as jump parameters to the MCMC. We found that the difference in BIC (Bayesian Information Criterion; \citealt{deWit2012}) does not favour the inclusion of a linear CV law in our model (Table~\ref{tab:MCMC_results}), with its $c_\mathrm{0}$ and $c_\mathrm{1}$ coefficients not significantly different from zero. Nonetheless it is noteworthy that the best-fit model corresponds to a decrease in CV from centre-to-limb by about 200\,m\,s$^{-1}$, which is of the same order as solar variations (on the 100s of m\,s$^{-1}$ level; \citealt{Dravins1982}). The inclusion of a quadratic CV law in the model yielded similar BIC value as the linear law (Table~\ref{tab:MCMC_results}), and we thus conclude that the data are most consistent with a constant CV across the stellar disk. 

\subsubsection{Differential rotation scenario} 
\label{sec:DR_scen} 
In a second step we considered a model with differential rotation (DR), using a solar-like DR law $\Omega(\theta_\mathrm{lat})$ = $\Omega_\mathrm{eq} (1-\alpha\, sin^{2}(\theta_\mathrm{lat}))$, where $\Omega$ is the rotation rate (equal to $\Omega_\mathrm{eq}$ at the equator), $\alpha$ is the relative differential rotation rate, and $\theta_\mathrm{lat}$ is the stellar latitude. Constraining the DR is one way to determine the exact regions of the star transited by the planet, and hence disentangle the true 3D spin-orbit geometry from projection effects. Indeed, in this case $v_\mathrm{eq}$ and $sin(i_{*})$ can be treated as independent parameters, as they influence differently the model stellar surface velocity (see Eq. 8 in \citealt{Cegla2016}). The jump parameters of the MCMC are thus $\alpha$, $\lambda$, $v_\mathrm{eq}$, and $cos(i_{*})$. We forbade $\alpha$ to go beyond 1 for physical reasons. We use $cos(i_{*})$ with a uniform prior in [-1 ; 1] so that $i_{*}$ can be bound directly to the range 0--180$^{\circ}$ (\citealt{Cegla2016}). We also used $sin(i_{*})$ as a prior to reflect the geometric probability of observing a given stellar inclination, but we note that removing this prior does not change significantly the discussion below. The results of the MCMC, displayed in Fig.~\ref{fig:mcmc_DR}, show asymmetric posterior probability distributions (PD) for all model parameters except $\lambda$. Therefore, in addition to the median and its associated 68.3\% confidence interval (Sect.~\ref{sec:method}), in Table~\ref{tab:MCMC_results} we also report the highest density intervals (HDI) that contains 68.3\% of the distribution mass, such that no point outside the interval has a higher density than any point within it\footnote{The HDI are equivalent to the range of values that encompass 68.3\% of the distribution on each side of the median when the distribution is Gaussian}. 

We compared our results with the measurements of $|\alpha_\mathrm{Kep}|$ obtained for Kepler G-type stars by \citet{Balona2016} (because they use photometric methods, only the absolute value of $\alpha_\mathrm{Kep}$ could be inferred). These authors derived an analytical formula to approximate $|\alpha_\mathrm{Kep}|$, which matches well their measurements for stars in the range $6000-6600$\,K (see their Eq.\,2 and Fig.\,11). Using this formula, we found that WASP-8 ($T_\mathrm{eff}$=6100\,K; $\Omega_\mathrm{eq}$ between 0.24 and 0.41\,rad\,d$^{-1}$, as derived from our HDI for $v_\mathrm{eq}$) would be expected to show relative differential rotation rates between 0.19 and 0.28. It is thus noteworthy that the PD of $\alpha$ and $v_\mathrm{eq}$ in our DR best-fit scenario (Fig.~\ref{fig:mcmc_DR}) display a common peak at about 0.25 and 2\,km\,s$^{-1}$, respectively.

\begin{figure}
\centering
\includegraphics[trim=0.75cm 2.8cm 0cm 0cm,clip=true,width=\columnwidth]{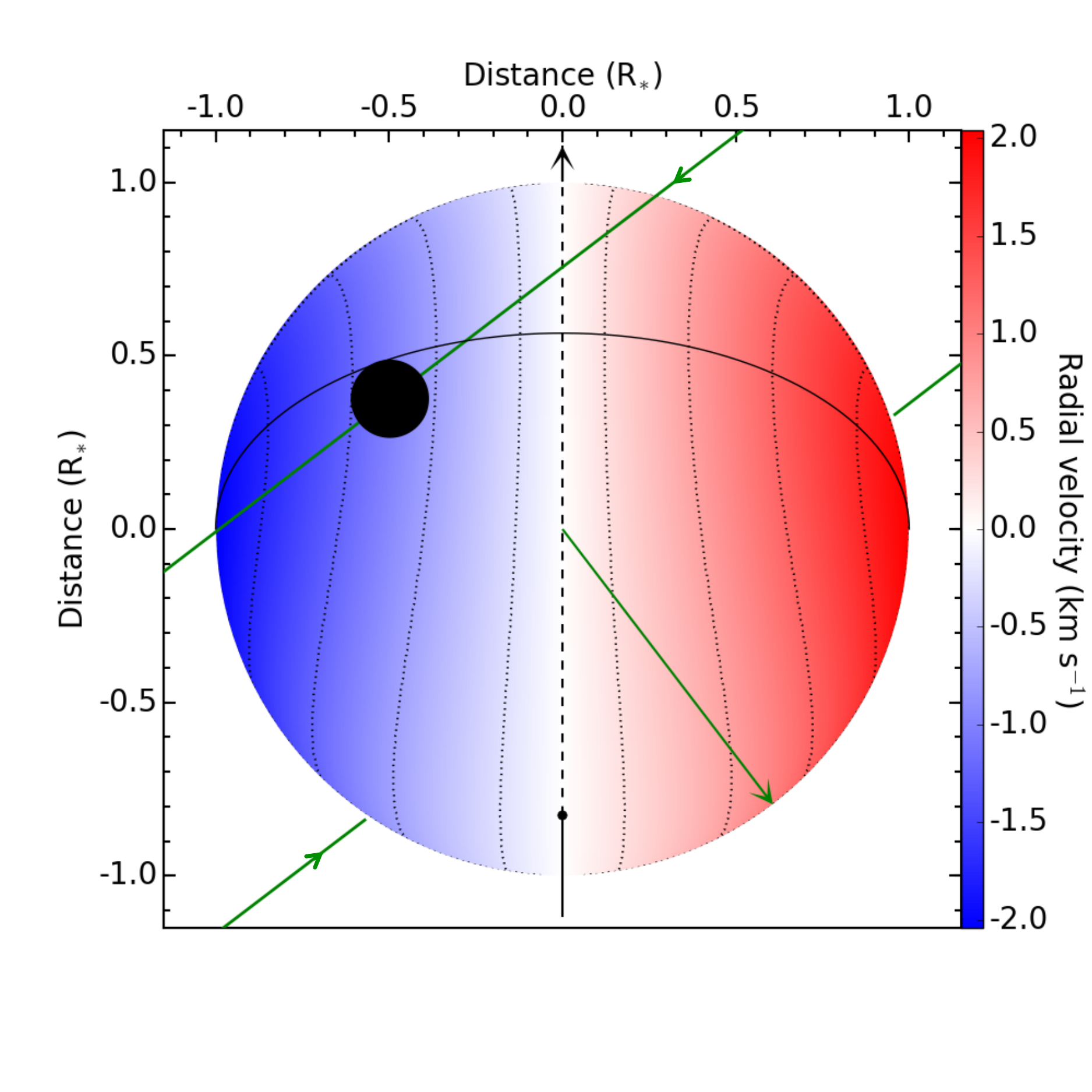}
\includegraphics[trim=-1.8cm 0cm 0.45cm 6cm,clip=true,width=\columnwidth]{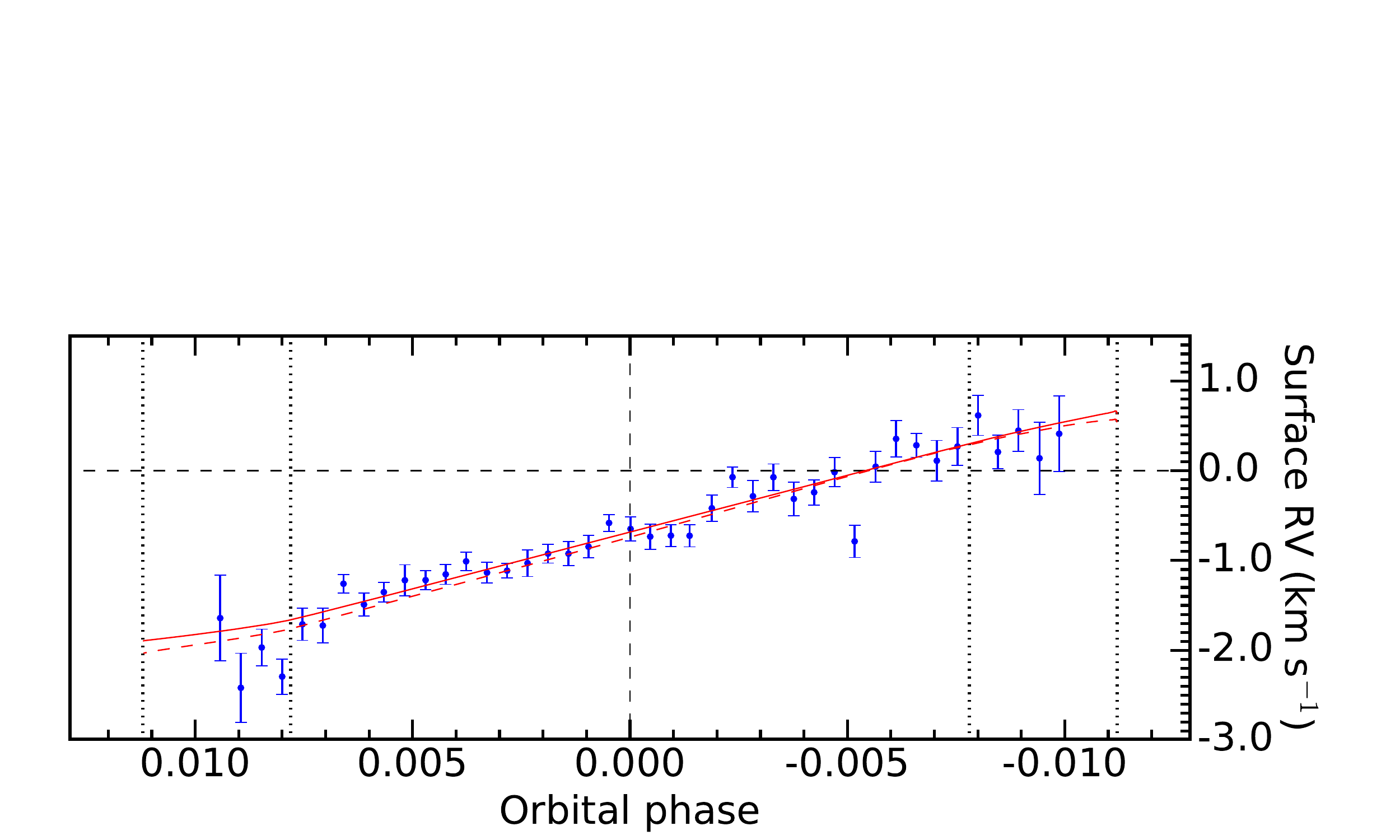}
\caption[]{\textit{Upper panel}: Projection of WASP-8 in the plane of sky, for the best-fit in the differential-rotation scenario. The stellar disk is colored as a function of its RV field, with thin black lines showing iso-RV curves equally spaced. The stellar spin axis is inclined and displayed as a black arrow going from the south pole to the north pole (dashed when within the star). The normal to the orbital plane is shown as a green arrow, and the orbital trajectory is displayed as a green curve. In the solid-body scenario, the spin axis orientation is degenerate and iso-RV curves would be parallel to the projected stellar spin. \textit{Lower panel}: Surface RVs of the regions occulted by the planet, as a function of orbital phase (increasing from right to left to match the retrograde motion of the planet). The measurements (blue points) are well reproduced in both the SB (solid red curve) and DR (dashed red curve) scenarios}
\label{fig:disque}
\end{figure}

Nonetheless a wide range of values for $\alpha$ and $v_\mathrm{eq}$ are allowed by the data. This is linked to the high stellar inclination obtained in the DR scenario, as one can see in Fig.~\ref{fig:disque}. The combination of the stellar north pole pointing away from the observer and a large sky-projected obliquity means the planet always occults regions that are either close to the projected stellar equator (where differential rotation is lowest) or are near the stellar spin axis (where RVs are low, and were measured with larger errors during ingress). As a result, RVs of the planet-occulted regions in the DR scenario are very similar to that of the SB scenario, even when $\alpha$ is large (see lower panel in Fig.~\ref{fig:disque}). Accordingly, the PD of $v_{eq} \sin i_{\star}$ derived from the equatorial velocity and stellar inclination in the DR scenario yields similar results as in the SB scenario (Table~\ref{tab:MCMC_results}). This is also the case for $\lambda$, because variations of the obliquity directly affect the stellar longitudes (and thus also the iso-RV curves) crossed by the planet, which cannot be compensated for by variations of the other parameters.  

Even though the best-fit for the DR scenario yields a slightly lower $\chi^2$ than in the SB scenario, the increase in BIC penalizes its additional free parameters (Table~\ref{tab:MCMC_results}). The same conclusion was reached using the Akaike information criterion. Therefore, we cannot conclude with certainty to the differential rotation of WASP-8 and we consider the solid-body model as the most plausible. We note that the iso-RV curves deviate more significantly from the SB case near the limbs of the star (Fig.~\ref{fig:disque}), and thus measurements of the surface RV at higher SNR during ingress/egress will be crucial in confirming differential rotation.


\begin{table*}[]
\caption[]{MCMC-derived properties of the WASP-8 system}
\centering
\begin{threeparttable}
\begin{tabular}{cccccccccc}
\hline
\hline
\noalign{\smallskip}
Scenario & $v_{eq}$ (km~s$^{-1}$) & $i{\star}$ ($^{\circ}$) & $\alpha$ & $\lambda$ ($^{\circ}$)  & c$_0$ (km~s$^{-1}$) & c$_1$ (km~s$^{-1}$) & c$_2$ (km~s$^{-1}$) & $\chi^2$ & BIC \\
\hline
\noalign{\smallskip}
SB & 1.90$\pm$0.05 & 90$^{\dagger}$ & 0$^{\dagger}$ & -143.0$\stackrel{+1.6}{_{-1.5}}$ & 0$^{\dagger}$ & 0$^{\dagger}$ & 0$^{\dagger}$ & 68.4 & 75.9 \\ 
\hline
\noalign{\smallskip}     
SB + CV & 1.90$\pm$0.05 & 90$^{\dagger}$ & 0$^{\dagger}$ & -143.2$\stackrel{+1.6}{_{-1.5}}$ & -0.16$\stackrel{+0.14}{_{-0.13}}$ & 0.22$\stackrel{+0.20}{_{-0.19}}$ & 0$^{\dagger}$ & 67.0 & 78.3 \\ 
\hline
\noalign{\smallskip}     
SB + CV & 2.02$\pm$0.09 & 90$^{\dagger}$ & 0$^{\dagger}$ & -139.8$\stackrel{+2.1}{_{-2.4}}$ & -1.06$\stackrel{+0.50}{_{-0.52}}$ & 3.68$\stackrel{+1.87}{_{-1.93}}$ & -2.81$\pm$1.51 & 63.3 & 78.3 \\ 
\hline
\noalign{\smallskip} 
DR & 2.5$\stackrel{+1.6}{_{-0.6}}$ & 124$\stackrel{+24}{_{-46}}$ & 0.3$\pm$0.5 & -142.6$\pm$1.9 & 0$^{\dagger}$ & 0$^{\dagger}$ & 0$^{\dagger}$ & 67.7 & 82.7 \\ 
 					 & [1.8,3.1] & [101,158] & [0.1,1.0] & [-144.5,-140.8] &  &  &  &  \\ 
\hline
\hline
\noalign{\smallskip}
  \end{tabular}
  \begin{tablenotes}[para,flushleft]
  Notes: In all scenarios, we report the median of the posterior distributions as the best estimate for the model parameters, along with its associated 68.3\% confidence interval. The $^{(\dagger)}$ symbol indicates fixed parameters. In the solid-body (SB) scenarios, the value in the $v_{eq}$ column corresponds to $v_{eq} \sin i_{\star}$. When CV are accounted for, c$_{0}$ is degenerate and must be derived from the posterior distribution of c$_{1}$ and c$_{2}$. It is thus not possible to fit a constant CV model, and in that case c$_{0}$ is set to 0\,km\,s$^{-1}$. In the differential-rotation (DR) scenario, some posterior distributions are asymmetric, and we also provide the 68.3\% HDI (see text).\\
  \end{tablenotes}
  \end{threeparttable}
\label{tab:MCMC_results}
\end{table*}

\section{Spatial variations of the stellar spectra}
\label{sec:shape_ccf}

\subsection{Analysis}

We analysed the variations in the shape of the local CCF$_\mathrm{loc}$ along the transit chord, by examining the FWHM and contrast of their Gaussian fits (Sect.~\ref{sec:reloaded RM}). We found that the contrast decreases almost linearly during the transit, from about 70\% at ingress down to about 45\% at egress (Fig.~\ref{fig:ctrst_sig}). We checked that varying the fixed system properties (Table~\ref{tab:fix}) by up to three times their uncertainty did not remove this trend. Performing the reloaded RM analysis with the master-out computed from CCF$_\mathrm{DI}$ taken before the transit only, or after the transit only, also did not remove the trend. Furthermore, these different master-out CCFs were found to be consistent between themselves (i.e., with differences in RV centroid, contrast, and FWHM lower than 1$\sigma$). The FWHM remains roughly constant while the planet is fully in front of the stellar disk, but increases by about 1\,km\,s$^{-1}$ at ingress and then again at egress, although the noise on the local FWHM is too large to confirm these variations. We also note that the fully in-transit CCF$_\mathrm{loc}$ have a lower FWHM than the master-out CCF$_\mathrm{DI}$ (7.253$\pm$0.02\,km\,s$^{-1}$, Fig.~\ref{fig:ctrst_sig}). This is likely due to the rotational broadening in the disk-integrated CCF. 

\subsection{Interpretation}

Instrumental effects, even the partial defocusing of the telescope during the first half of the transit (Sect.~\ref{sec:data_red}), are unlikely to explain such a continuous and large variation in the measured contrast. The reloaded RM technique is sensitive to the light curve used to normalize the CFFs. In particular, a larger transit depth can decrease the contrast of the extracted CCF$_\mathrm{loc}$. However, we found that to obtain a constant local contrast during the transit of WASP-8b, the light curve would have to be abnormally asymmetric. Indeed, by comparison with the Mandel \& Agol light curve (\citealt{Mandel2002}) used by \citet{queloz2010}, the required light curve would have to be deeper by more than 1\% during ingress and the first part of the transit, and shallower by up to 0.5\% during the last part of the transit. Such variations are excluded by the photometric measurements obtained by \citet{queloz2010}. Furthermore, these variations spread over the whole transit would likely arise from wide bands of spots and plages strongly varying in contrast with stellar latitude, which would be inconsistent with the assumption of a constant local contrast from the planet-occulted regions. Finally, we note that even the required asymmetric light curve would not change significantly the RV time series, except during ingress and egress where the deviations to the Mandel \& Agol light curve would be the largest. Thus, we find it more likely that the measured contrast variation traces variations in the depth of the lines emitted by the stellar photosphere, arising, for example, from the effect of the stellar magnetic field. Using theoretical line profiles generated in the frame of a previous study (\citealt{Cegla2016}, albeit with a G2V spectral type in the MHD simulations), we investigated whether changes in the magnetic field strength could yield the observed contrast variations. However, even for the Fe\,I-617.3\,nm line, particularly sensitive to the magnetic field, an increase in field strength from 20 to 500\,Gauss only changed the contrast by a few percent, much lower than the measured 25\%. 

\begin{figure}
\centering
\includegraphics[trim=1.5cm 1.75cm 1cm 2cm,clip=true,width=\columnwidth]{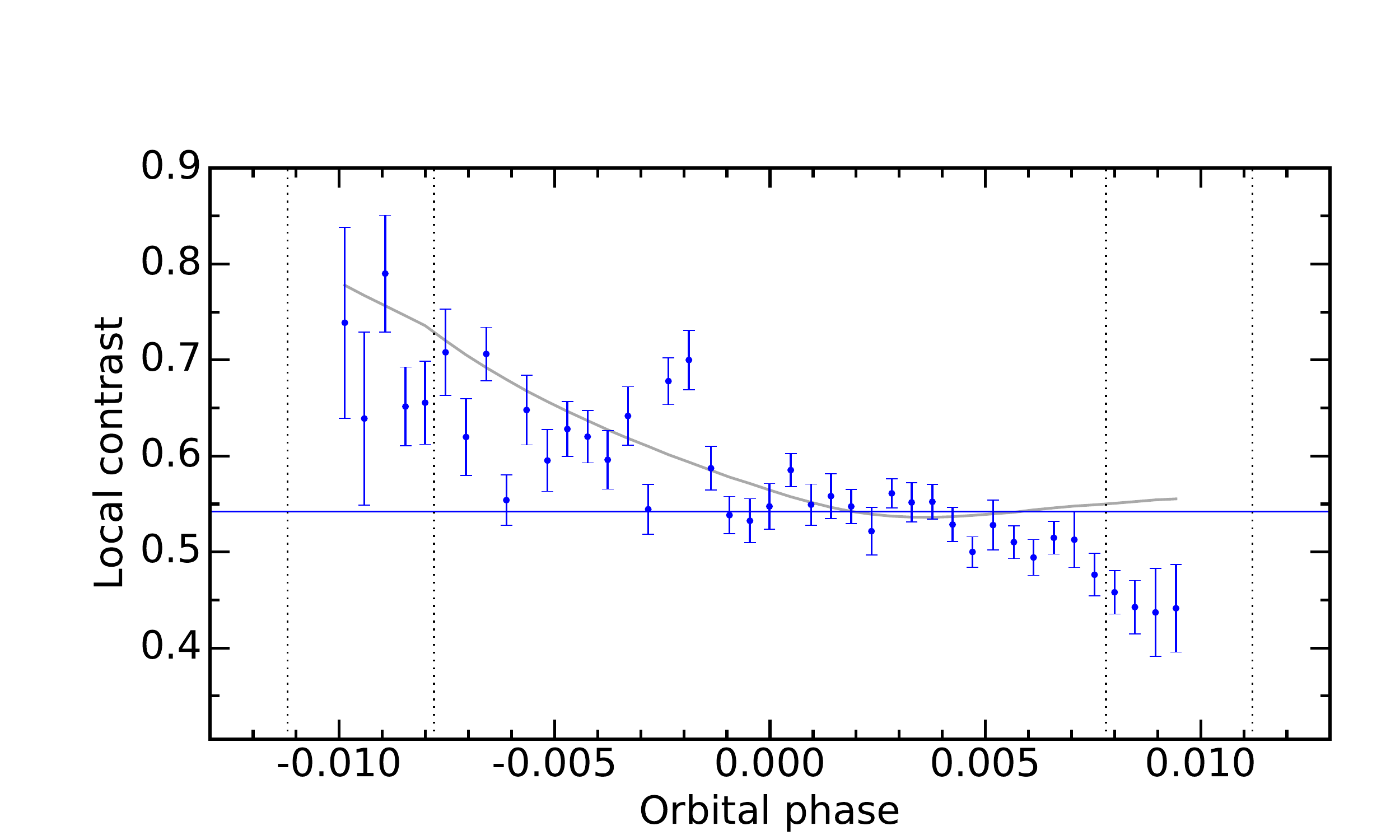}
\includegraphics[trim=1.5cm 0cm 1cm 2.7cm,clip=true,width=\columnwidth]{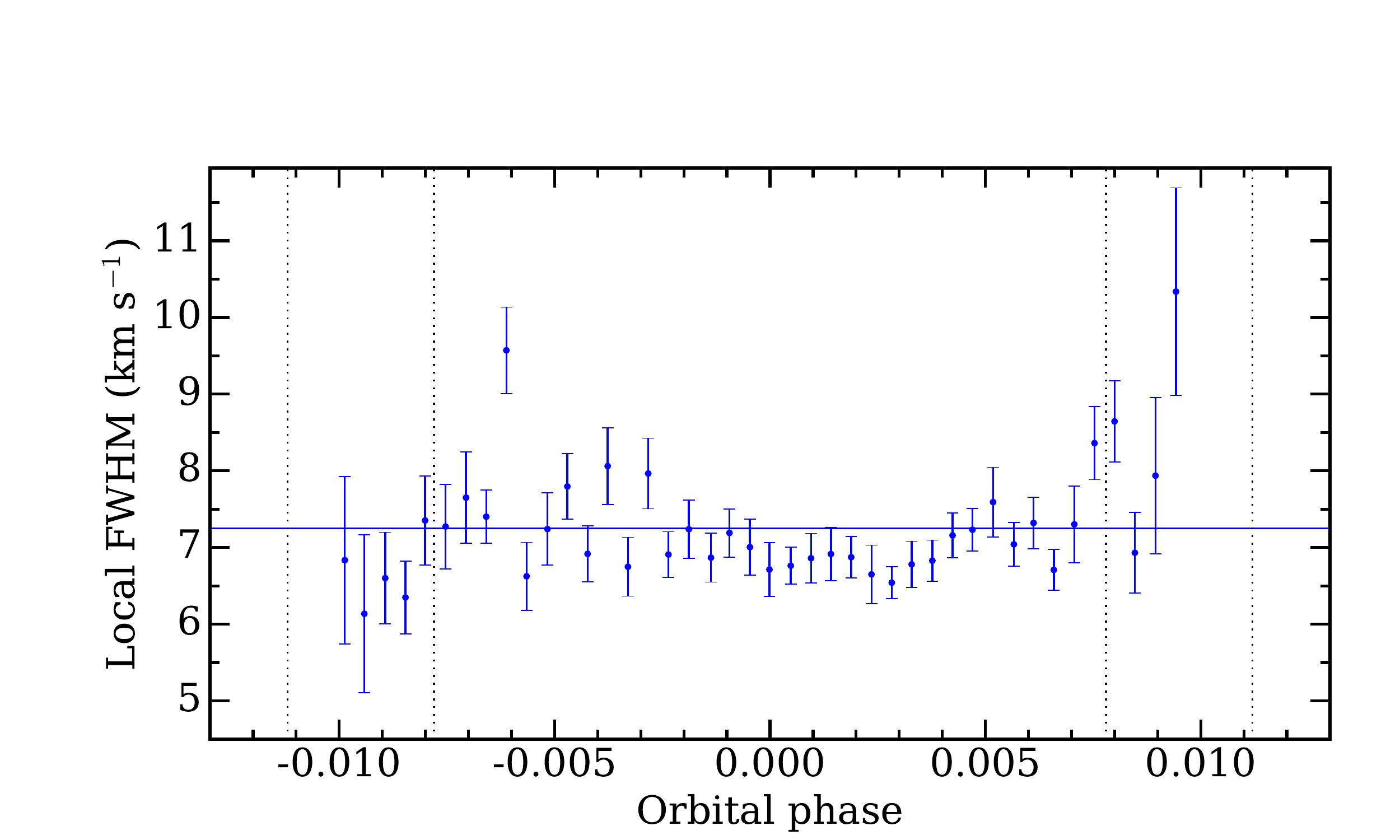}
\caption[]{Contrast (\textit{upper panel}) and FWHM (\textit{lower panel}) of the local CCF$_\mathrm{loc}$ from the planet-occulted regions, as a function of orbital phase. A significant decrease in contrast occurs along the transit chord, while the FHWM remains stable overall. The solid horizontal lines correspond to the contrast and FWHM of the disk-integrated CCF$_\mathrm{DI}$ measured outside of the planet transit. Vertical dotted lines indicate the planet contacts. The grey line in the upper panel shows our best fit for the local contrast, assuming it varies linearly from the pole to the equator of the star.}
\label{fig:ctrst_sig}
\end{figure}

The high obliquity of the system suggests that WASP-8b crosses many stellar latitudes along the transit chord (Fig.~\ref{fig:disque}). Following an empirical approach, we thus considered a simple model where the contrast of the CCF$_\mathrm{loc}$ decrease linearly from the poles to the equator of the star, and the stellar inclination is allowed to vary such that $v_{eq} \sin i_{\star}$ = 1.9\,km\,s$^{-1}$ (following the SB scenario, Table~\ref{tab:MCMC_results}). We then simulated the transit of WASP-8b, and calculated the contrast of the average local profiles from planet-occulted regions. We used $\chi^2$ statistics to compare the measured and simulated contrast series, varying the polar and equatorial contrasts, and the stellar inclination, in our model. We also included in the fit the comparison between the contrasts of the out-of-transit model CCF$_\mathrm{DI}$ and the observed master-out CCF$_\mathrm{DI}$ (Fig.~\ref{fig:ctrst_sig}). The best fit was obtained for a polar contrast of 100\%, an equatorial contrast of 52\%, and a stellar inclination i$_{\star}$=112\,$^{\circ}$. The corresponding theoretical contrast along the transit chord is displayed in Fig.~\ref{fig:ctrst_sig}. Interestingly, the inclination we derive falls within the range obtained from the independent analysis of the surface RV (Sect.~\ref{sec:DR_scen}). In that configuration, WASP-8b would cross the equator of the star at near the center of the transit. Incidentally, this is also the time when the local contrast becomes similar to the contrast of the master-out CCF$_\mathrm{DI}$ (0.5423$\pm$9$\times$10$^{-4}$, Fig.~\ref{fig:ctrst_sig}), which might indicate that the equatorial bands of the stars contribute the most to the shape of the disk-integrated CCF.

We note that a polar contrast of 100\% is likely unphysical. However this value is a projection of our gradient model at the pole, and the variation in contrast effectively observed and fitted actually goes from about 80\% to 50\% during the first part of the transit. We also note that our gradient model does not explain the continuous decrease of the measured contrast during the last part of the transit. To yield a linear decrease in contrast along the full transit chord the simulated star would have to point toward the observer, so that the planet always occult the same stellar hemisphere. Yet this is inconsistent with the stellar inclination derived from the RV analysis, which corresponds to the star pointing away from the observer. Furthermore, with the star pointing toward us we found that the out-of-transit contrast was overestimated because the star is then mostly visible through its polar regions, which have the deepest line contrast. In conclusion, the measured contrast can be explained during most of the transit if the star points away from the observer and its local contrast decreases linearly at high stellar latitudes. The situation is more complex at the end of the transit, when the planet occults the equatorial bands of the star where the local CCFs are both shallower and broader (Fig.~\ref{fig:ctrst_sig}). Further investigations, beyond the scope of this paper, will be required to investigate whether Zeeman splitting or photospheric temperature changes could explain these variations.

\section{Pitfalls of the velocimetric and tomographic RM analysis}
\label{sec:pitfalls}

Neglecting chromatic effects, the occultation of the stellar disk by the transiting planet decreases the flux in every part of the disk-integrated CCF$_\mathrm{DI}$ by the same percentage. In terms of absolute flux, the decrease is thus smaller at the bottom of the stellar lines than in their continuum, resulting in an apparent (positive) ``bump" in the in-transit CCF$_\mathrm{DI}$ that is maximized at the velocity of the planet-occulted region. When the planet transits the stellar hemisphere that is rotating towards the observer, the occulted regions are blueshifted and the bump is located in the blue wing of the CCF$_\mathrm{DI}$. The fit to this distorted CCF$_\mathrm{DI}$ with a Gaussian profile thus yields a RV centroid that is redshifted with respect to the Keplerian RV of the star (blueshifted when the planet transits the stellar hemisphere rotating away from the observer). The amplitude of this anomalous shift depends on the amplitude of the bump, i.e. on both the planet-to-star radius ratio, and the contrast and width of the intrinsic local CCF behind the planet. 

Both the velocimetric and tomographic analysis of the RM effect assume a constant shape for the local CCF$_\mathrm{loc}$ across the stellar disk (e.g., \citealt{ohta2005}, \citealt{cameron2010a}), yet the case of WASP-8b show that the local contrast in particular can vary significantly along the transit chord (Sect.~\ref{sec:shape_ccf}). An increase in the CCF$_\mathrm{loc}$ contrast increases the amplitude of the planet bump in the CCF$_\mathrm{DI}$, resulting in a larger RV deviation to the Keplerian curve (and vice versa for a contrast decrease). In a velocimetric analysis of the RM effect, the planet will thus appear to be occulting regions farther away from the projected stellar spin axis than it is in reality. In the case of WASP-8b, the planet transits the redshifted half of the stellar disk for a short time before occulting its blueshifted half (Sect.~\ref{sec:res_velfield}), which results in a small blueshifted RV anomaly at the beginning of the transit followed by a larger redshifted RV anomaly, as detected by \citet{queloz2010}. Yet, because the contrast of the CCF$_\mathrm{loc}$ decreases steadily during the transit (Fig.~\ref{fig:ctrst_sig}), we argue that the planet is closer to the spin axis at the beginning of the transit, and farther away toward the end of the transit, than derived from the velocimetric analysis of \citet{queloz2010}. If correct, this would have led \cite{queloz2010} to underestimate the absolute value of the sky-projected obliquity, which is in agreement with our measurement of $\lambda$ = -143.0$\stackrel{+1.6}{_{-1.5}}^{\circ}$ larger by more than 4$\sigma$ from the \citet{queloz2010} value (-123.0$\stackrel{+3.4}{_{-4.4}}^{\circ}$). We also argue that the variation in local contrast biased their RM-derived estimation of $v_{eq} \sin i_{\star}$ (1.59$\stackrel{+0.08}{_{-0.09}}$\,km\,s$^{-1}$) by $\sim$3$\sigma$. It is interesting to note that our value for $v_{eq} \sin i_{\star}$ = 1.90$\pm$0.05\,km\,s$^{-1}$ is consistent with their spectroscopic value (2.0$\pm$0.6\,km\,s$^{-1}$) derived from the fit of the stellar Fe\,I lines in the HARPS spectra. In conclusion, variations in the shape of the local CCF$_\mathrm{loc}$ cannot be detected in the traditional velocimetric analysis of the RM effect, despite clearly leading to strong biases in the derived properties. 

Since Doppler tomography compares the profiles of the CCF$_\mathrm{DI}$, rather than their RV centroid, it should be less impacted than a traditional velocimetric analysis regarding the determination of the planet trajectory and $v_{eq} \sin i_{\star}$. This is because the signature of the planet is directly included in the tomographic model of the stellar line fitted to the in-transit CCF$_\mathrm{DI}$. Therefore, variations in the contrast and FWHM of the local CCF$_\mathrm{loc}$ should not significantly affect the localization of the planet signature. However the usual assumption of a constant shape for the CCF$_\mathrm{loc}$ would likely bias both the estimation of the planet-to-star radius ratio and the width of the intrinsic stellar+instrumental profile. 

Alternatively, the reloaded RM technique allows for a simple and efficient way to isolate the CCF$_\mathrm{loc}$ and assess its variations. The residuals between the CCF$_\mathrm{loc}$ of WASP-8 and their fits do not show significant dispersion or spurious features (Fig.~\ref{fig:CCF_locales}), giving us confidence that the shape of the CCF$_\mathrm{loc}$ is well approximated by a Gaussian profile along the transit chord. As such, a variation in the $CCF_{loc}$ contrast or FWHM does not change the RV centroid, and therefore does not affect our retrieval of the stellar surface RVs.


\section{Conclusion}
\label{sec:conclu}

Analysis of the RM effect during the transit of an exoplanet allows us to measure the dynamical properties of the stellar photosphere, and to constrain the system architecture through the measurement of the spin-orbit alignment. While the velocimetric and tomographic analysis of the RM effect yield the sky-projected obliquity of a system, the reloaded RM method developed by \citet{Cegla2016} can potentially constrain the differential rotation of the host star. In that case, breaking the $v_{eq} \sin i_{\star}$ degeneracy also yields the measurement of the stellar inclination and true 3D obliquity. Even when the reloaded RM technique is unable to constrain differential rotation, it avoids some of the biases that can affect the more traditional techniques. Indeed, our application of the reloaded RM technique to the WASP-8 system revealed near-linear variations of up to 35\% in the contrast of the local line profile along the transit chord. Those variations could not have been detected in the velocimetric analysis of the RM effect performed by \citet{queloz2010}, and led to a significant misestimation of the sky-projected obliquity and projected stellar rotational velocity of the system. Such biases in the measured obliquity, caused by changes in the local CCFs shape, were predicted by \citet{Cegla2016a}. Since the reloaded RM approach relies on the analysis of the RV centroid of the local CCF$_\mathrm{loc}$, it is not sensitive to variations in their FWHM and contrast as long as they keeps a symmetric shape. This is the case for WASP-8, for which the local line profile is well approximated by a Gaussian, and we thus refine the properties of the system to $\lambda$ = -143.0$\stackrel{+1.6}{_{-1.5}}^{\circ}$ and $v_{eq} \sin i_{\star}$ = 1.90$\pm$0.05\,km\,s$^{-1}$. We note that our study does not change the conclusions from \citet{queloz2010} that WASP-8b is on a strongly misaligned, retrograde orbit, nor does it change the subsequent interpretations in the literature.

We did not find significant convective velocity variations from the center to the limb of WASP-8. We also found no definitive evidence for differential rotation, although the posterior distribution for $\alpha$ in this scenario peaks at about 0.25, consistently with the measurements obtained by \citet{Balona2016} for Kepler stars similar to WASP-8. The constraints obtained on $\alpha$ partially lift the degeneracy between $v_{eq}$ and $i_{*}$, suggesting possible values in the range 1.8-3.1\,km\,s$^{-1}$ and 101-158$^{\circ}$. The distributions derived for $i_{*}$ and $\lambda$ would also yield the true 3D obliquity (see calculation in \citealt{Cegla2016}) in the range 106-129$^{\circ}$. Even though measurements obtained with higher accuracy during the ingress/egress of WASP-8b will be necessary to confirm or infirm the differential rotation of its host star, we emphasize that the posterior distribution for the sky-projected obliquity is well defined and yields the same results whether differential rotation is accounted for or not.
 
In conclusion, we caution against using a velocimetric analysis to interpret the RM effect of transiting planets, as it is sensitive to variations in the shape of the local stellar CCF$_\mathrm{loc}$ that cannot be detected except through a direct study of the spectra. Doppler tomography should be less affected by such variations, although they could still lead to biases in the estimation of the planet-to-star radius ratio and the width of the stellar line profile. It is possible that the properties of many planetary systems analyzed in the literature values have been biased by variations in the shape of the local CCF$_\mathrm{loc}$. The reloaded RM-derived stellar surface velocities can still be biased by variations in the symetry of the stellar line profile, or deviations to the light curve used to normalize the CCF$_{DI}$ (caused, e.g., by spots). Nonetheless, this method allows for a simple and efficient way to isolate the local stellar CCF$_\mathrm{loc}$, assess their variability, and interpret correctly their RV centroid even in case of variations in their contrast or FWHM. Furthermore, because the reloaded RM technique extracts the average stellar properties behind the occulting planet surface, small Earth-size planets that will be detected by missions like CHEOPS, TESS and PLATO will act as finer probes of the stellar surface in terms of spatial resolution. Although the lower flux retrieved from behind the surface of small planets will make it more challenging, analyzing the RM effect with this technique will thus have a strong potential when coupled with observations from future generations spectrographs like ESPRESSO, SPIROU or NIRPS. \\


\begin{acknowledgements}
We warmly thank the referee for their appreciative reading of our paper. This work has been carried out in the frame of the National Centre for Competence in Research ``PlanetS'' supported by the Swiss National Science Foundation (SNSF). V.B. and A.W. acknowledges the financial support of the SNSF.
\end{acknowledgements}

\bibliographystyle{aa} 
\bibliography{biblio} 

\begin{appendix}

\section{Additional Posterior Probability Distributions}
\label{apn:mcmc}

\begin{center}
\begin{figure*}[b!]
\centering
\includegraphics[trim=0cm 13cm 6.75cm 2.5cm, clip, scale=1.]{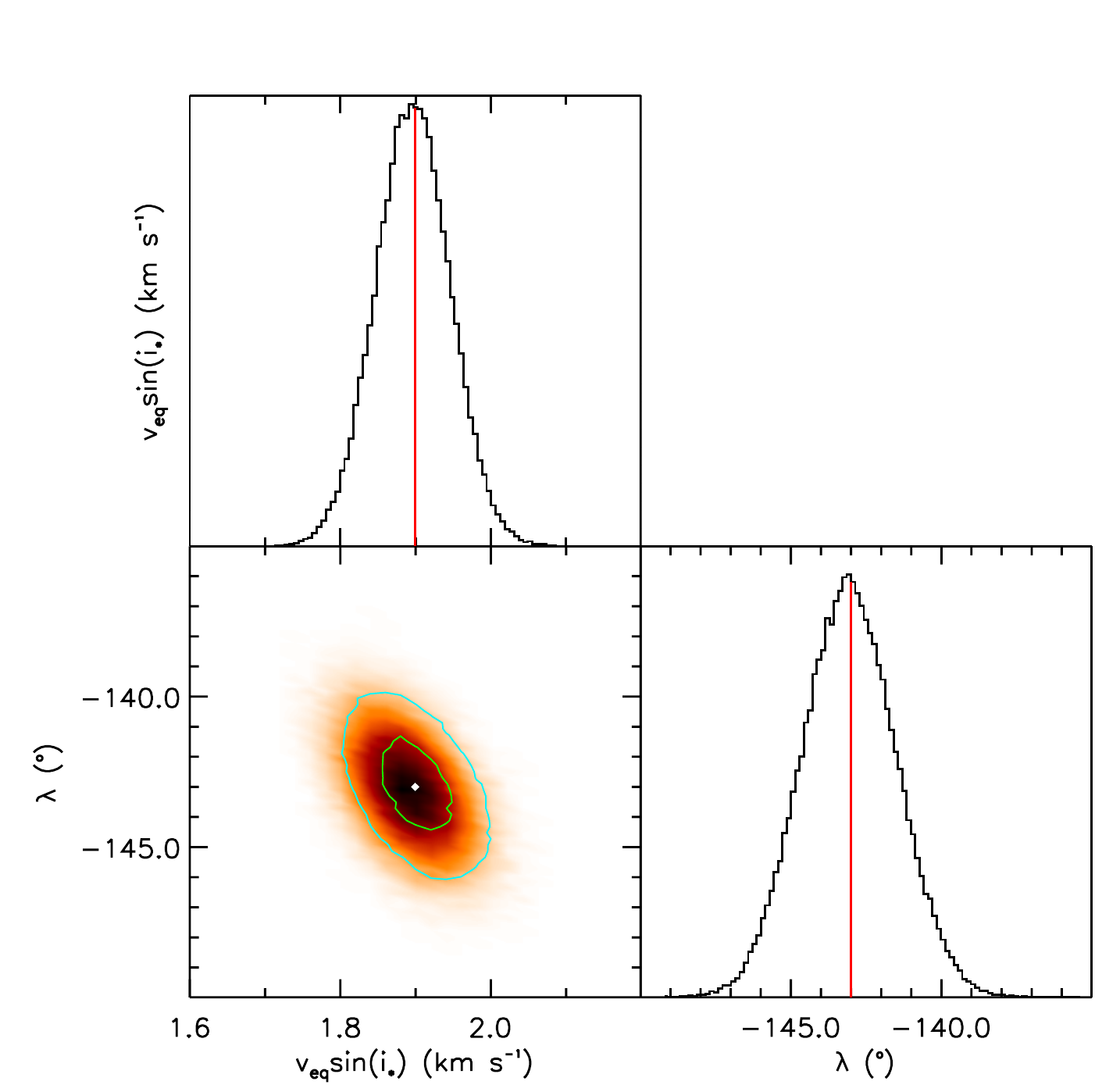}
\caption[]{Correlation diagrams for the probability distributions of the solid-body rotation model parameters. Green and blue lines show the 1 and 2$\sigma$ simultaneous 2D confidence regions that contain respectively 39.3\% and 86.5\% of the accepted steps. 1D histograms correspond to the distributions projected on the space of each line parameter. The red line and white point show median values.}
\label{fig:mcmc_solidbody}
\end{figure*}
\end{center}

\begin{center}
\begin{figure*}[b!]
\centering
\includegraphics[trim=0cm 13cm 4cm 3cm, clip,scale=1.2]{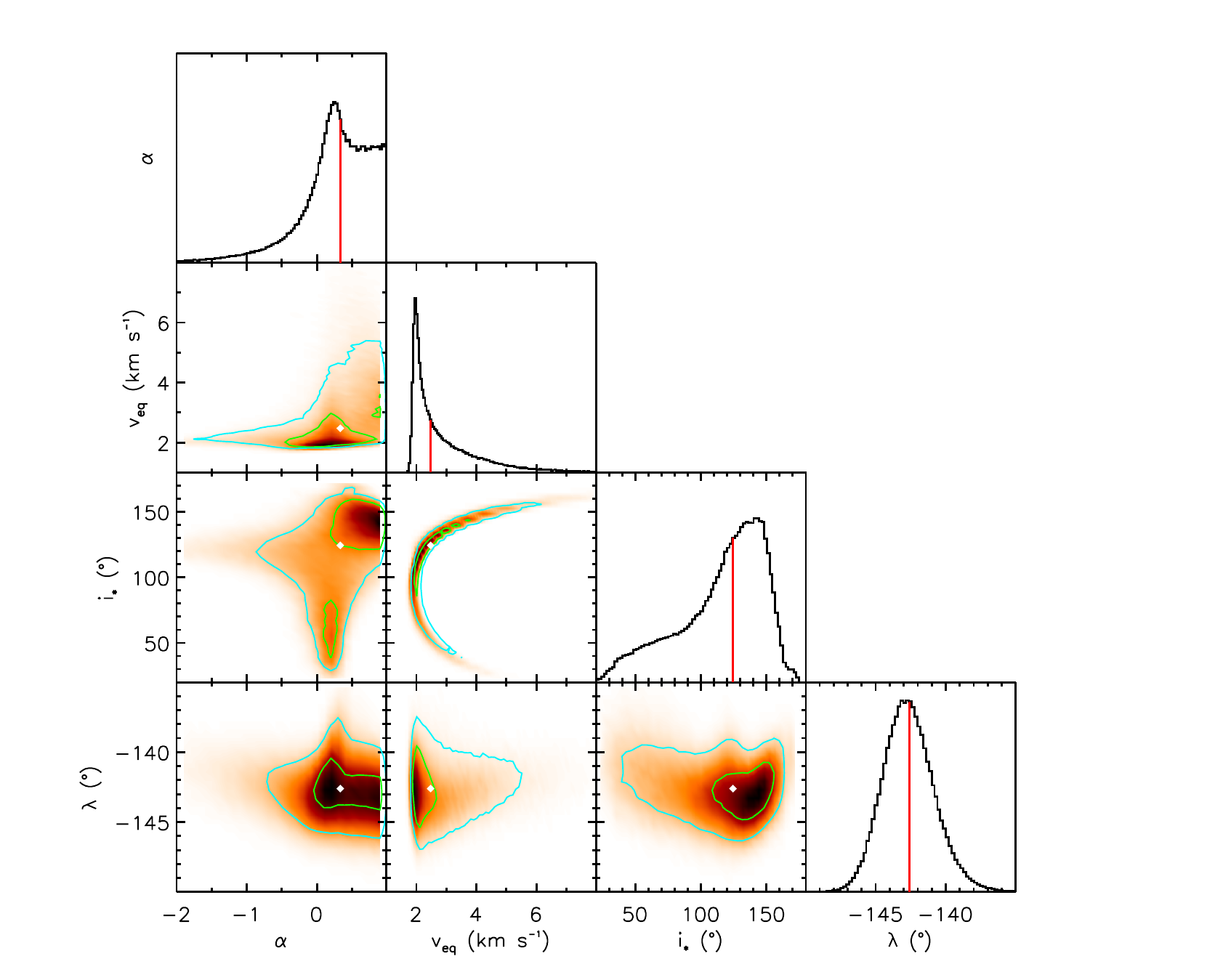}
\caption[]{Correlation diagrams for the probability distributions of the differential rotation model parameters. Green and blue lines show the 1 and 2$\sigma$ simultaneous 2D confidence regions that contain respectively 39.3\% and 86.5\% of the accepted steps. 1D histograms correspond to the distributions projected on the space of each line parameter. The red line and white point show median values. } 
\label{fig:mcmc_DR}
\end{figure*}
\end{center}

\end{appendix}

\end{document}